\documentclass[journal]{IEEEtran}
\usepackage{cite}
\usepackage{float}\usepackage{amssymb}
\usepackage{amsfonts}
\usepackage{booktabs}
\usepackage[pdftex]{graphicx}
\usepackage[cmex10]{amsmath}
\usepackage{algorithm}
\usepackage{algorithmic}
\usepackage{array}
\usepackage[caption=false,font=footnotesize]{subfig}
\usepackage{colortbl}
\usepackage{float}
\usepackage{subfig}
\usepackage{amsmath}
\usepackage{enumerate}
\usepackage{cite}
\usepackage[american,english]{babel}
\usepackage{xspace}
\usepackage{tabularx}
\usepackage{lipsum}
\usepackage{balance}

\usepackage{setspace}
\usepackage{amsmath}

\usepackage{tikz}

\usepackage[justification=centering]{caption}

\definecolor{Gray}{gray}{0.9}

\usepackage{amsmath,amsfonts,amssymb,amsthm}
\usepackage{mathrsfs}
\usepackage{hyperref}

\usepackage{caption}
\usepackage{graphicx}
\usepackage{subcaption}
\usepackage{empheq}
\usepackage{diagbox} 
\usepackage{multirow}
\usepackage{tabularx}
\usepackage{xcolor}  
\usepackage{pifont}  

\begin{document}
\title{Large Language Model (LLM)-enabled Reinforcement Learning for Wireless Network Optimization}	
\author{Jie~Zheng, Ruichen~Zhang, Dusit Niyato, Haijun~Zhang, Jiacheng~Wang, Hongyang~Du,  Jiawen~Kang, Zehui~Xiong

\thanks{J. Zheng is with State-Province Joint Engineering and Research Center of Advanced Networking and Intelligent Information Services, College of Computer Science, Northwest University,  Xi'an, 710127, Shaanxi, China. (jzheng@nwu.edu.cn) (Corresponding author: Jie Zheng.)}
\thanks{R. Zhang, D. Niyato and J. Wang are with the College of Computing and Data Science, Nanyang Technological University, Singapore 639798. (ruichen.zhang@ntu.edu.sg, dniyato@ntu.edu.sg, jiacheng.wang@ntu.edu.sg)}
\thanks{H. Zhang is with the Institute of Artificial Intelligence, University of Science and Technology Beijing, Beijing 100083, China. (zhanghaijun@ustb.edu.cn)}
\thanks{H. Du is with Department of Electrical and Electronic Engineering, the University of Hong Kong, China. (duhy@eee.hku.hk)}
\thanks{J. Kang with the Automation of School, Guangdong University of Technology, Guangzhou 510006, China.(kavinkang@gdut.edu.cn)}
\thanks{Z. Xiong is with the Queen's University Belfast, BT7 1NN, United Kingdom (zehuixiong@sutd.edu.sg)}
}
\maketitle

\begin{abstract}
Enhancing future wireless networks presents a significant challenge for networking systems due to diverse user demands and the emergence of 6G technology. While reinforcement learning (RL) is a powerful framework, it often encounters difficulties with high-dimensional state spaces and complex environments, leading to substantial computational demands, distributed intelligence, and potentially inconsistent outcomes. Large language models (LLMs), with their extensive pretrained knowledge and advanced reasoning capabilities, offer promising tools to enhance RL in optimizing 6G wireless networks. We explore RL models augmented by LLMs, emphasizing their roles and the potential benefits of their synergy in wireless network optimization. We then examine LLM-enabled RL across various protocol layers: physical, data link, network, transport, and application layers. 
Additionally, we propose an LLM-assisted state representation and semantic extraction to enhance the multi-agent reinforcement learning (MARL) framework. This approach is applied to service migration and request routing, as well as topology graph generation in unmanned aerial vehicle (UAV)-satellite networks.
Through case studies, we demonstrate that our framework effectively performs optimization of wireless network. Finally, we outline prospective research directions for LLM-enabled RL in wireless network optimization.


\end{abstract}

\section{Introduction}
With the advent of sixth-generation (6G) networks, the wireless communication landscape is undergoing transformative changes, marked by increasing complexity holographic wireless network and a broader range of user requirements~\cite{10819473}. 
Thus, advanced optimization techniques have become indispensable for harnessing the full capabilities of 6G systems. 

Among these technological innovations, reinforcement learning (RL) emerges as a critical enabler~\cite{9372298}. 
RL demonstrates intrinsic suitability for addressing dynamic challenges in wireless networks due to its adaptive decision-making and self-optimizing properties. By engaging in iterative environmental interactions through trial-and-error mechanisms, RL agents autonomously optimize predefined reward functions across operational trajectories. 
Additionally,  with a wealth of knowledge and expertise, large language models (LLMs) are one of the most promising candidates for 6G network~\cite{10685369}. 

\begin{figure}[!t]
	\centering
	\includegraphics[width=0.48\textwidth]{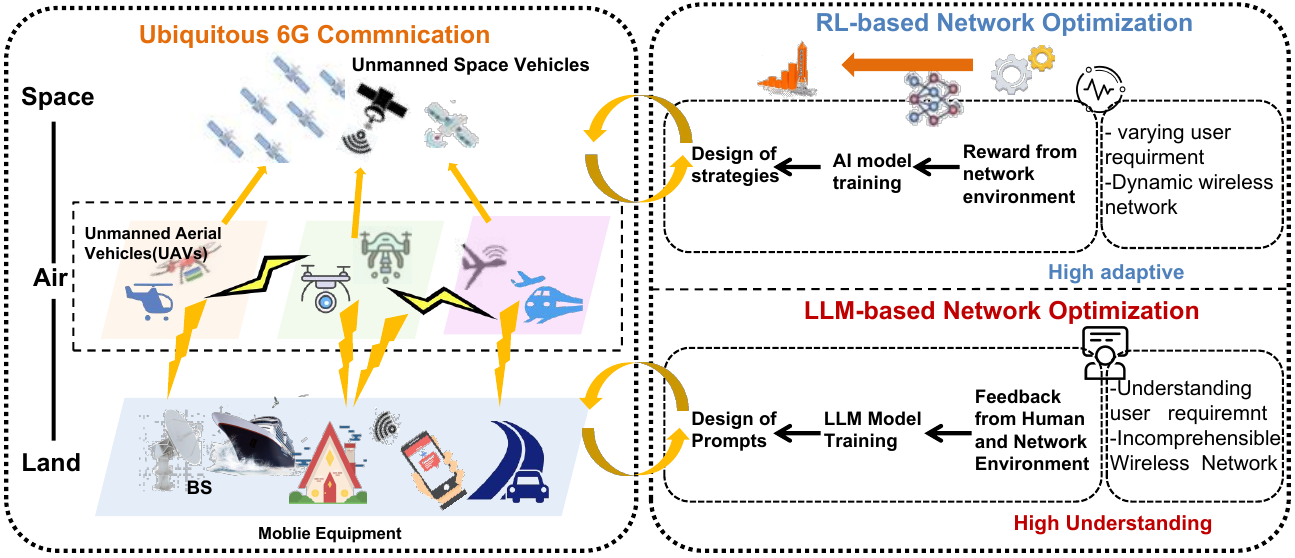}%
	\caption{Motivation for integrating LLMs with RL for wireless network optimization. The yellow arrows represent the 6G network provides its current state with LLMs and a subsequent reward signal to the RL agent.}
	\vspace{-6mm} 
\label{fig:motivation}
\end{figure}

LLMs can be used to assist RL agents, leading to better performance in multitask learning, greater sample efficiency, and improved high-level task planning~\cite{10766898}. As illustrated in Figure~\ref{fig:motivation}, combining the unique strengths of RL and LLMs for the joint design of wireless networks is a promising approach. However, this integration also raises a critical question:
\begin{itemize}
    \item \emph{How to integrate LLMs in classical agent–environment interaction of RL in wireless networks?}
\end{itemize}

By integrating the strong representation and reasoning abilities of LLMs with the adaptive decision-making of RL, large model-driven RL can more effectively understand complex wireless environments, such as dynamic channels, multi-user interference, and resource constraints along with human or expert knowledge. This enables more efficient exploration and learning, improving policy performance and generalization. Their integration helps manage high-dimensional network states, interpret high-level objectives such as maximizing throughput or minimizing latency, and translate them into actionable RL strategies such as dynamic power allocation and interference management, overcoming the limitations of traditional RL in wireless network optimization. Although LLMs-enabled RL presents potential benefits for network management, several significant questions emerge:

\begin{itemize}
    \item \textbf{Q1:} What kind of LLM-enabled RL paradigm can offer to the design of wireless optimization?
    \item \textbf{Q2:} How to apply LLMs-enabled RL in wireless network optimization?
\end{itemize}

To answer these questions, this paper addresses the key questions discussed above. The key contributions are outlined below.
\begin{itemize}
    \item \textbf{First:} We explore the integration of the emerging field of integrating LLM into the RL paradigm in wireless network.  We present a framework to systematically classify LLMs within the traditional agent-environment paradigm, as feature extractor, reward designer, policy interpreter, decision-maker, incorporating these aspects into the RL paradigm for optimizing wireless networks. 
    
    \item \textbf{Second:} We investigate challenges and solutions associated with the integration of LLMs-enabled RL for wireless network and examine the different issues in different protocol layers: physical layer, data link layer, network layer, transport layer, application layer.

    \item \textbf{Third:} We develop a novel LLMs-enabled multi-agent reinforcement learning (MARL) for service migration and request routing with generated graph representation and semantic extraction in unmanned aerial vehicle (UAV)-satellite networks. Experimental results validate the effectiveness of our proposed framework. Furthermore, we provide a potential research directions.

\end{itemize}

\begin{figure}[!t]
	\centering
	\includegraphics[width=0.40\textwidth]{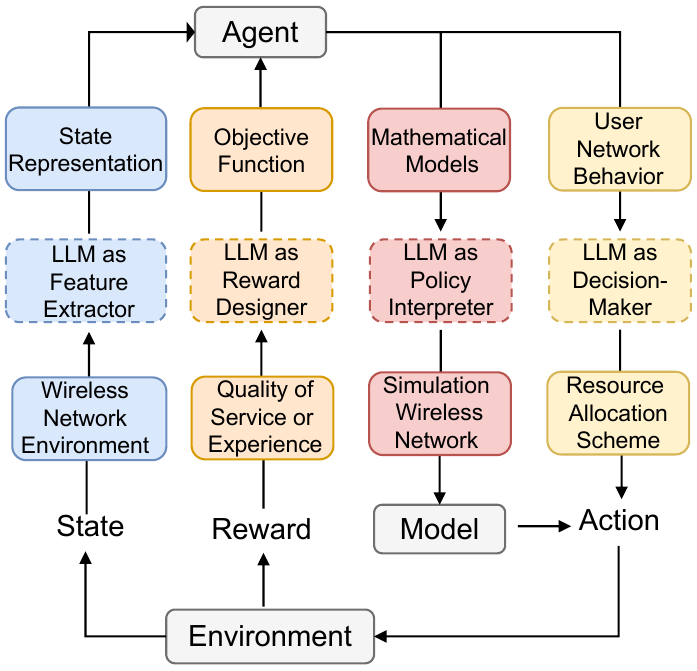}%
	\caption{LLMs-enabled RL for wireless network with four roles.} 
	\vspace{-4mm} 
\label{fig:framework}
\end{figure}

\section{The concepts of LLM-enabled RL}

\begin{figure*}[!t]
	\centering
	\includegraphics[width=0.8\textwidth]{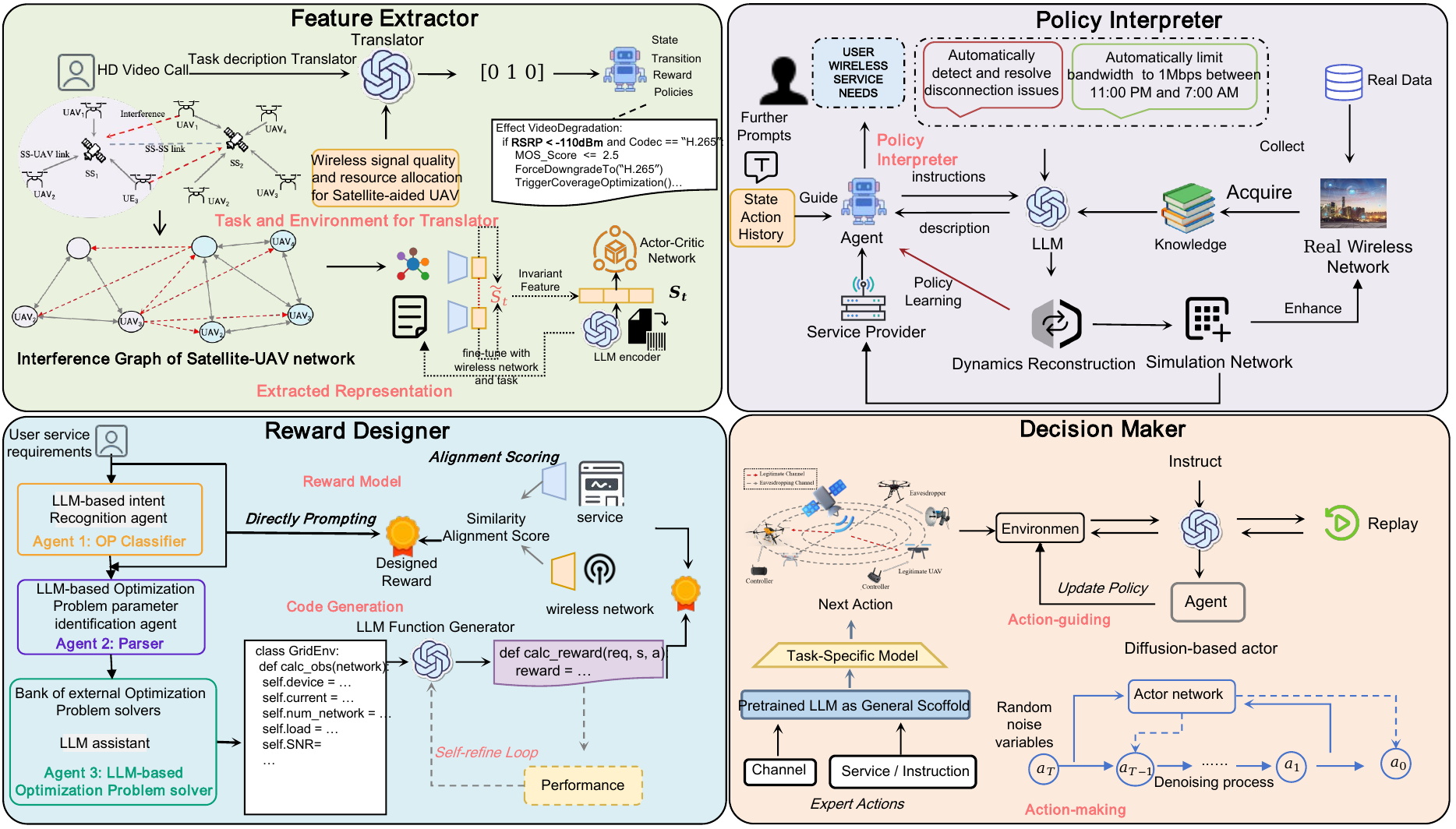}%
	\caption{Summary of typical LLMs-enabled RL for wireless network with LLMs as feature extractor, reward designer, policy interpreter, decision-maker.} 
	\vspace{-6mm} 
\label{fig:paradigm}
\end{figure*}

\quad In this section, we analyze LLMs' potential in enhancing the capabilities of RL. Then, we discuss the diverse roles of LLMs within the RL framework for optimizing wireless networks.

\subsection{Overview of LLM and RL}
\quad LLMs play a crucial and impactful part in tackling the persistent difficulties of optimizing wireless networks via RL. The subsequent content outlines the essentials of LLMs and RL.

\begin{itemize}
    \item \textbf{RL:} In conventional RL, an agent learns through environment interaction, using trial and error to maximize rewards~\cite{9372298}. Actions are chosen based on the current state, with the goal of optimizing the policy for cumulative rewards, typically defined as a Markov decision process (MDP). It consists of four key elements: the state space, action space, transition probability, and reward function. RL agents make precise decisions using diverse information, such as robots executing natural language commands or visual games with language instructions. Traditional RL struggles with interpreting complex data and adapting in dynamic settings, facing challenges such as low sample efficiency, designing reward functions for multimodal data, and ensuring robust generalization across tasks.

    \item \textbf{LLMs-enabled RL:} LLMs offer solutions with their ability to understand natural language and process visual data. This enables LLMs to interpret and respond to complex multimodal information, enhancing RL applications. LLM-enabled RL incorporates LLM capabilities in multimodal processing and reasoning, improving RL by using knowledge-rich LLMs~\cite{10766898}. 
    In contrast to pretrained models that are unable to learn from their surroundings, LLM-enabled RL acquires task-specific data through interactions with the environment, thereby supporting specialization and dynamic adaptation via continual learning. 
    For instance, an RL agent guiding a robotic arm learns to pick and place objects in a warehouse. The LLM helps interpret natural language instructions (“stack the blue boxes on the left”) and extract relevant visual or sensor features as the robot interacts with different objects and places them accordingly.

\end{itemize}

Thus, to improve efficiency of RL in wireless networks, it is essential to design novel RL algorithms, with particular emphasis on those empowered by LLMs.

\subsection{Overview of LLM-enabled RL for wireless network}
LLMs enhance RL in wireless networks by serving in four key roles, as illustrated in Figure~\ref{fig:framework}. First, LLMs can act as feature extractor for RL, helping to represent the state of the wireless network environment. Second, LLMs can serve as reward designers by defining the objective function from quality of service (QoS) or quality of experience (QoE). Third, LLMs function as policy interpreters for RL, enabling the simulation of real wireless networks using real data and mathematical models. Finally, LLMs can act as decision makers in RL, supporting resource allocation schemes based on user and network behaviors. Their synergy forms a closed-loop optimization cycle of perception–definition–simulation–execution, thereby enhancing the adaptability and interpretability of RL in wireless networks. We then describe the four roles in detail, and provide an illustrative example, as shown in Figure~\ref{fig:paradigm}.

\begin{itemize}

     \item \textbf{Feature Extractor:} When tasks in wireless networks involve linguistic or visual components, RL agents often face challenges in understanding and optimizing control strategies. This is particularly evident in scenarios where network management requires interpreting natural language instructions or visual data, such as satellite imagery or network topology maps. These types of data introduce complexity that makes it difficult for RL agents to directly map them to actionable network optimization strategies. An LLM can significantly aid in overcoming these challenges by processing multimodal data, extracting key features, and converting natural language into task-specific languages that the RL agent can understand. The world models focus on learning dynamic state transitions, while LLMs can interpret the meaning of these states and user requirements expressed in natural language.

\emph{Example:} By processing raw image data from embodied AI agents, LLMs can extract key semantic features to enhance the accuracy of environmental representations to benefit vehicular communication by enabling a data reduction of over 90\%~\cite{zhang2025embodied}.

     \item \textbf{Reward Designer:} In complex tasks with sparse rewards or challenging reward functions, LLMs utilize their knowledge, reasoning, and code skills in two ways: as an implicit reward model offering context-based reward via training or prompting, and as an explicit reward creating code for reward functions, clarifying the logic of reward computation based on environmental parameters and tasks. 
     Furthermore, LLMs can collaborate with Large Reasoning Models to formulate complex, multi-objective rewards that reflect logical trade-offs and constraints, going beyond simple numerical optimization. LLMs interpret user QoS requirements and enhance reward functions for wireless models, balancing accuracy and communication efficiency while optimizing service quality and resource use. 

\emph{Example:} 
RL agents leveraging LLM-designed rewards demonstrate consistently superior performance, achieving up to 6.2\% lower energy consumption with a 2.0 Mbits packet size compared to counterparts using hand-crafted rewards in low-altitude economic network (LAENet)~\cite{cai2025large}.

\begin{small}
\begin{table*}[t!]
\centering
\caption{Summary of LLM-enabled RL for different layers}
\begin{tabular}{p{1.2cm}|p{1.1cm}|p{3.3cm}|p{8.0cm}}
\hline
\rowcolor{gray!10}{\footnotesize \textbf{Layers}} & \footnotesize \textbf{Reference} & \footnotesize \textbf{Technique}  & \footnotesize \textbf{Pro \& Cons}  \\
\hline
\rowcolor{blue!5}\footnotesize \textbf{Physical Layer} & \footnotesize \cite{10812008}  & \scriptsize   
\textbullet{} Channel representation \newline
\textbullet{} Energy-efficient reward \newline
\textbullet{} Electromagnetic map generator\newline
\textbullet{} Resource allocation actions
    &  \scriptsize 
     \textcolor{green}{\ding{51}}  Transformer encoding and prompt of channel state information, \newline
     \textcolor{green}{\ding{51}}  Capacity, energy-efficient and signal-to-noise evaluator \newline
    \textcolor{red}{\ding{56}}  Time-varying channel, high overhead, and representation mismatch, reward misaligned and interpretability \newline 
    \textcolor{red}{\ding{56}}  compute-intensive, online update and high-dimensional action  
    \\
\hline
\rowcolor{pink!5}\footnotesize \textbf{Data Link Layer} & \footnotesize \cite{10839354} &  \scriptsize 
\textbullet{} Link quality representation \newline
\textbullet{} Average packet loss reward \newline
\textbullet{} Generated backoff behavior \newline
\textbullet{} Link selection actions
& \scriptsize 
     \textcolor{green}{\ding{51}}  LLM-based rate model, Packet-loss penalty rule \newline
     \textcolor{green}{\ding{51}}  LLM-driven MAC contention
     markov backoff models,traffic-burst GANs \newline
    \textcolor{red}{\ding{56}}  Different link protocol,User mobility, Collision detection \newline 
    \textcolor{red}{\ding{56}}  Combines delay, packet-loss
    into scalar reward via LLM  
  \\
\hline

\rowcolor{green!5} \footnotesize\textbf{Network Layer} & \footnotesize \cite{10742906} & \scriptsize
\textbullet{} Connectivity expression\newline
\textbullet{} End-to-end delay rewards \newline
\textbullet{} Generated topology \newline
\textbullet{} Load balanced actions
& \scriptsize 
     \textcolor{green}{\ding{51}}  Topology generated graph
Dynamic topology rule, \newline
     \textcolor{green}{\ding{51}}  Maps delay throughput overhead into one reward via LLM \newline
     \textcolor{green}{\ding{51}} Chooses routing paths or protocol parameters in dynamic graphs \newline
    \textcolor{red}{\ding{56}}  Dynamic topology, traffic varying
graph size explosion, partial observability \newline 
    \textcolor{red}{\ding{56}}  Combinatorial explosion of
route options, balance traffic, routing and loading 
   \\
\hline

\rowcolor{blue!5} \footnotesize\textbf{Transport
Layer} & \footnotesize \cite{10720863} & \scriptsize
\textbullet{} Transport process expression \newline
\textbullet{} Jitter reward\newline
\textbullet{} Virtual transport scenarios \newline
\textbullet{} Congestion control actions   
& \scriptsize 
     \textcolor{green}{\ding{51}}  Transformer encoding congestion window and RTT sequences \newline
     \textcolor{green}{\ding{51}} Generated flow
mixes, burst-pattern simulation, captures congestion dynamics \newline
    \textcolor{red}{\ding{56}}  Non-stationary network
conditions; unstable reward dynamics \newline 
    \textcolor{red}{\ding{56}}  Poor match to real network
transport process
   \\
\hline

\rowcolor{pink!5} \footnotesize\textbf{Application Layer} & \footnotesize \cite{10819462} & \scriptsize
\textbullet{} User intent expression \newline
\textbullet{} Qos and QoE reward \newline
\textbullet{} Chain of service generator \newline
\textbullet{} Service slicing action
& \scriptsize 
     \textcolor{green}{\ding{51}}  GPT-crafted user-request
scenarios, Transformer embedding of QoS logs \newline
     \textcolor{green}{\ding{51}}  RL-tuned Qos,  QoE and service level agreement (SLA) rewards via LLM \newline
     \textcolor{green}{\ding{51}} LLM-generated flow
mixes,burst-pattern simulation for RL \newline
    \textcolor{red}{\ding{56}}  User behavior bias from intent; \newline 
    \textcolor{red}{\ding{56}}  Conflicting objectives; noisy
user feedback
   \\
\hline

\end{tabular}
\vspace{-6mm}
\end{table*}
\end{small}

    \item \textbf{Policy Interpreter:} 
Model-based RL relies on accurate models for policy learning in wireless networks, where interpretability is crucial. LLMs provide two critical functions: as an environment simulator and a policy explainer. As a simulator, an LLM analyzes historical network logs and channel fading to create a high-fidelity virtual testbed that predicts future states and rewards for the RL agent. As a policy explainer for XRL, it translates the agent’s decisions into natural language, clarifying the reasoning, such as link and channel selection.

    \emph{Example:} To address the “black-box” issue of DRL in 6G network slicing, the XRL framework uses LLMs to make logic transparent decisions, thereby improving trust and understanding in wireless resource management~\cite{10588921}.

    \item \textbf{Decision-Maker:} LLMs offer promising solutions for RL in two ways: action-making and action-guiding, for wireless network. 
    
    In \emph{action-making}, offline RL is approached by LLMs as a form of sequence modeling, where rewards are utilized to guide the generation of actions. Their extensive pre-training improves learning speed. LLMs as decision-makers convert task instructions into quantitative constraint boundaries, creating a wireless optimization space. 
    
    \emph{Example:} Lightweight LLMs, via prompt engineering, generate a high-quality action space to streamline RL-based task scheduling for mobile edge intelligence~\cite{10819462}. 

    In \emph{action-guiding}, 
    This expert knowledge such as Telecom-LLMs~\cite{10685369}, when used in policy learning, boosts sample efficiency by transferring to the wireless RL agent.

    \emph{Example:} By processing high-level operator intents, a lightweight LLM transforms ambiguous goals into a validated, machine-readable objective, thereby establishing a precise target for a hierarchical RL agent by optimally initiating network applications that improve 12\% throughput, 17.1\% energy efficiency, and 26.5\% delay~\cite{10978505}.

\end{itemize}

Thus, RL enhanced by LLMs for optimizing wireless networks focuses on multi-objective tasks across different network types with various roles. It manages multi-modal data together with natural language instructions and designs detailed reward systems by incorporating metrics from various fields relevant to wireless networks.

\section{LLM-enabled RL for networking:different issues in different protocol layers}

\quad In this section, we analyze issues and solutions and examine the challenges related to the implementation of LLMs-enabled RL in different protocol layers.

\begin{figure*}[!t]
	\centering
	\includegraphics[width=0.8\textwidth]{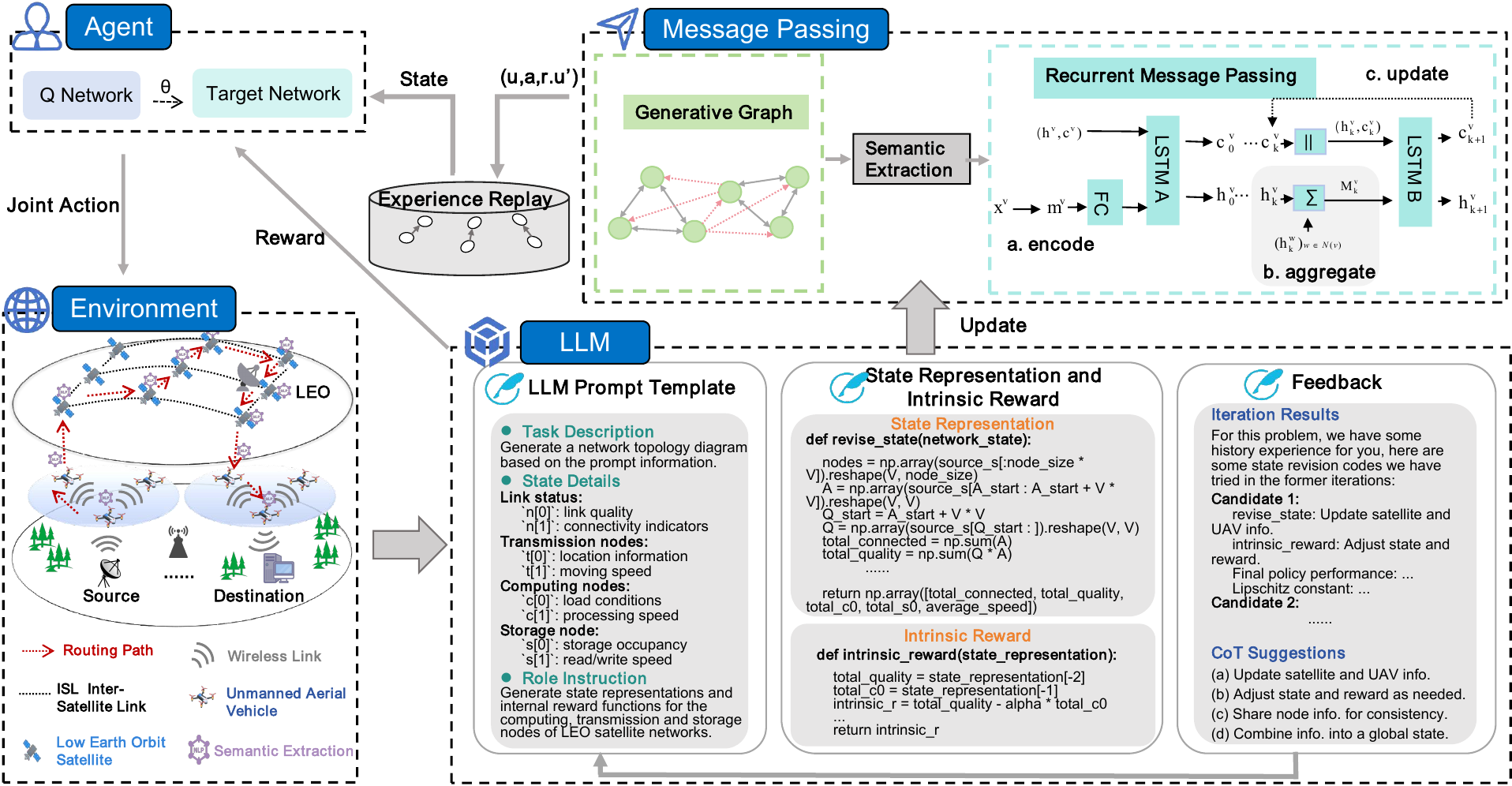}%
	\caption{LLMs-enabled RL for service migration, consisting of four function: (1) UAV-satellite network environment; (2) LLM-enabled RL with prompt template, state representation and intrinsic reward, and feedback; (3) Message Passing: message-passed observation from the recurrent model; (4) RL agent with deep Q network.} 
	\vspace{-6mm}
\label{fig:case}
\end{figure*}

\subsection{Research Issues and Solutions}
\quad As LLMs are integrated into RL in wireless network,  the decision-making capabilities of system and context-aware optimization of network can be enhanced. However, certain challenges still exist across various protocol layers as shown in Table I.

\begin{enumerate}

\item \textbf{LLMs-enabled RL for Physical Layer Optimization:}
Traditional RL in the physical layer struggles to model the complex spatial dependencies and context in wireless electromagnetic maps, which describe signal propagation, path loss, and interference. Thus, its adaptability is limited, making it hard to allocate wireless resources effectively to meet user requirements.
However, LLMs excel at processing contextual information and transferring semantic knowledge, making them uniquely suited to enhance RL frameworks. An RL agent enhanced with LLMs for joint communication and control design (JCCD) in 6G can efficiently integrate both radio frequency (RF) and non-RF modalities for interpreting wireless networked control systems~\cite{10812008}. 
The contextual reasoning of LLMs enables agents to interpret representations of channel dynamics and interference behavior, leading to more informed decisions for beamforming, power control, modulation schemes, and spectrum allocation in physical layer. Simulation results indicate that LLMs-enabled RL for JCCD achieves significantly outperforming the conservative Q-learning scheme trained with  25\% expert datasets.

\item \textbf{LLMs-enabled RL for Data Link Layer Optimization:}
LLMs can significantly enhance the RL agent’s capability by accurately forecasting link quality, user movement. AutoHMA-LLM integrates LLMs with traditional RL algorithms to tackle the challenges of task coordination and link scheduling in complex and dynamic wireless environments~\cite{10839354}.
This foresight enables the RL agent to proactively adjust link selection policies, contention resolution mechanisms, and channel allocations, leading to substantial improvements in link layer efficiency and fairness. 
Moreover, LLMs’ ability to interpret evolving link specifications and emerging data link patterns, particularly those associated with 6G requirements and real-time intelligent content services, ensures that RL agents can promptly adapt their decision-making strategies in response to updates in link configuration and rate demands. 
AutoHMA-LLM improved task completion accuracy by 5.7\%, reduced the number of communication steps by 46\%, and lowered token consumption by 31\% compared to baseline methods.

\item \textbf{LLMs-enabled RL for Network Layer Optimization:}
Traditional medium access control (MAC) protocols, optimized for static or low-mobility networks. These limitations manifest as frequent access update, escalated latency, and diminished throughput, especially in scenarios involving rapid topology changes, such as satellite-terrestrial networks or UAV network. For example, data-driven MAC protocols has been addressed in~\cite{10742906}, including task-oriented neural protocols constructed using MARL, and language-oriented semantic protocols harnessing large language models and generative models. 
Experimental results show that language-oriented semantic protocols increase average goodput by up to 1.95 times compared to retrained task-oriented protocols, and up to 2.45 times over S-ALOHA.

\item \textbf{LLMs-enabled RL for Transport Layer Optimization:}
LLMs exhibit exceptional capabilities in processing variable-context sequential data.  In~\cite{10720863}, LLMs-aided multi-agent transport are employed to extract semantic information for multi-agent communication,  and utilize their generative abilities to predict future actions, allowing agents to make more informed and effective decisions. 
Transport layers involve multidimensional QoS metrics (e.g., latency-reliability tradeoffs), where parameter priorities shift with application demands. The transformer architecture in LLMs autonomously recalibrates feature importance across protocol hierarchies, enabling agents to prioritize mission-critical constraints for intelligent content delivery. Experimental results show that LLMs-assisted multi-agent transport streamlines the transport process and decreases transport overhead by about 53\% relative to baseline methods.

\item \textbf{LLMs-enabled RL for Application Layer Optimization:}
For application-layer optimization, RL approaches face the challenge of capturing the semantic diversity inherent with diversity user services. 
LLMs demonstrate exceptional proficiency in interpreting contextual semantics across variable-length service requirements. 
An innovative RL framework enhanced by a lightweight LLM is proposed for adaptive application-layer control in multi-cloud task scheduling~\cite{10819462}. This approach leverages the inference capabilities of lightweight LLMs to assist the RL agent by generating potential action candidates, thereby accelerating policy exploration and improving overall scheduling performance. The integration of LLMs serves to enrich the agent’s decision-making process, enabling more efficient handling of independent task scheduling within complex multi-cloud environments.
Experimental results show that the use of a lightweight LLM to improve task scheduling decisions in RL leads to a 5.46\% reduction in costs compared to the more cost-effective genetic algorithm.

\end{enumerate}

\subsection{Lesson Learned}
Integrating RL with LLMs aims to develop refined cross-layer optimization strategies that address end-to-end task optimization and improve overall network efficiency. Importantly, LLMs will be used to establish end-to-end mappings from application-layer intents (e.g., smooth video playback) to physical-layer parameters (e.g., beamforming), thereby guiding RL toward globally optimal solutions.

\section{LLM-enabled MARL framework for service migration in wireless network}

In this section, we present an LLM-enabled MARL framework for service migration, combined with request routing and dynamic graph generation. Then, we validate the framework’s effectiveness in optimizing computational offloading within UAV-satellite integrated networks.

\subsection{Proposed Framework}
We integrate an LLM-enabled state representation (LESR)~\cite{wang2024LESR} into an MARL framework for UAV–satellite
network graph. As shown in Figure~\ref{fig:case}, the proposed framework contains the prompt description, state representation and intrinsic reward, and feedback as follows. 

\textbf{Prompt Description:} Task Description: The RL environment focuses on a satellite-UAV network, where the primary objective is to optimize service migration and request routing between UAVs and satellites. This optimization aims to achieve a balance between high network performance and low user costs. State Details: The state information includes details related to each dimension of the source state for each satellite-UAV node within each time slot. Role Instruction: LLMs is required to generate a task-relevant state representation by constructing a connected topological graph from the raw node data in python code format. With the feedback, we use LLM to improve the graph representation of satellite-UAV network to enhance the performance of service migration.

\textbf{State representation and intrinsic reward:}
To provide the LLM with both local data and a more holistic (though approximated) network view, enabling it to synthesize a comprehensive understanding for service migration decisions. This crucial part combines raw local observations with the refined, approximated global graph observations provided by the recurrent message-passing model. As detailed in Figure~\ref{fig:case}, LLMs are utilized to aid in state representation and the generation of intrinsic rewards. The cyclic message-passing mechanism enables distributed agents to iteratively propagate and integrate multi-hop spatiotemporal information, supporting enhanced decentralized decision-making.

We propose an LLM-assisted reward function to balance network quality and cost, formulated as $reward = (\alpha_1 \times throughput -\alpha_2 \times latency) \times penalty$, where $\alpha_1$ and $\alpha_2$ are weighting factors. Throughput represents the number of packets successfully delivered to their destination per time step. This serves as the “quality” metric in our reward.
Latency measures the time delay for packets to travel from source to destination of service migration. This acts as the “cost” metric. 

\textbf{Policy interpretation and Decision Maker:}
We leverage the LLM for semantic extraction of local topological information in wireless networks. For example, when satellite node A detects a change in the local topology, the original updated information can be represented as structured data: "nodeid": "A", "neighbors": ["id": "B", "status": "active", "bandwidth": "80Mbps", "latency": "5ms", "id": "id": "C", "status": "inactive", "bandwidth": "0Mbps", "latency": "N/A"]. To efficiently convey critical updates, we prompt the LLM to summarize this topology change in natural language, focusing on major events such as failures or changes in key metrics. Then, the LLM may output: ”A: Links B(80Mbps/5ms), C disconnected.”  Upon receiving this compressed message, adjacent nodes can input it back into the LLM for reverse parsing, reconstructing the original structured actions, such as: expand this topology update into JSON format with bandwidth/latency/disconnect flags.

\textbf{Feedback:}
This feedback mechanism allows the LLM to refine its state representations over time, adapting to the dynamic and imbalanced nature of the network load. By incorporating past experiences, LESR ensures that the generated state representations remain relevant and effective in capturing the evolving network conditions.

\subsection{Evaluation}
\begin{figure}[!t]
	\centering
	\includegraphics[width=0.45\textwidth]{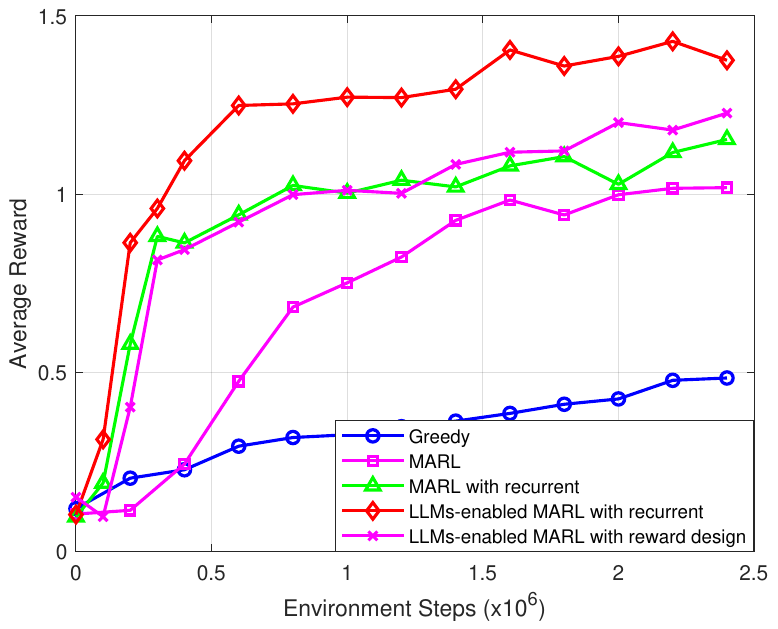}%
	\caption{Performance evaluation of our proposed LLMs-enabled MARL with recurrent framework. } 
	\vspace{-6mm}
\label{fig:results}
\end{figure}

To evaluate our proposed solution, we established a simulated Low Earth Orbit (LEO) satellite network comprising 20 satellites in 4 polar orbits (5 per orbit) at a 500 km altitude and 98° inclination. Each satellite has between 2 and 6 variable links, with intra-orbit bandwidths of 20–80 Mbps and inter-orbit bandwidths of 1–10 Mbps. Data packets range from 100 to 1000 Kb in size. Additionally, UAVs maintain dynamic, bidirectional connections to in-range satellites with link speeds of 0.1–0.8 Mbps~\cite{10375570}. The RL agent interacts with the environment, represented as a satellite-UAV graph. The graph state is processed by a GNN that combines non-recursive and recurrent layers to capture both static and temporal features. The agent employs a Deep Q-Network (DQN) for policy learning. The DQN’s policy head is a single linear layer with an input dimension of 256 from the GNN’s embedding.

Figure~\ref{fig:results} demonstrates that, once LLMs are integrated into the RL framework, the average reward rapidly converges to a level significantly higher than that of other service migration methods. The greedy optimization method, essentially a traditional shortest-path algorithm, performs the worst overall. While it may be effective in static graphs, it fails to handle the high variability typical of satellite network environments. In contrast, the recurrent approach employs a network structure that dynamically adapts to graph changes. When MARL enhanced with state representation and semantic extraction of LLMs, compared with LLMs-enabled MARL with reward design, our approach surpasses both the original recurrent and acyclic models in terms of convergence speed. This suggests that LLM-based MARL agents to learn optimal migration policies more efficiently, resulting in improved about 25\% decision-making performance gain for service migration in UAV-satellite scenarios.

\section{PROSPECTS FOR THE FUTURE\label{sec:future}}
This section outlines fundamental prospects and future pathways for RL powered by LLMs in  wireless networks.

\begin{itemize}

    \item \textbf{Robust and Security}: 
    We will incorporate wireless domain knowledge and protocol rules through knowledge graphs or rule-based constraints to guide LLMs away from unrealistic recommendations. An adaptive confidence evaluator will dynamically assess the trustworthiness of LLM outputs, allowing RL to adjust its reliance on these suggestions and preventing failures caused by low-confidence predictions. Federated LLM-RL can be discussed for privacy-preserving distributed optimization.



    \item \textbf{Integration of Wireless World Models and Large Reasoning Models}: Future research should build frameworks that tightly integrate LLMs with specialized wireless world model and large reasoning model, enabling interference environmental understanding, stronger planning and inference, and more robust end-to-end wireless network optimization. A key direction is exploring how LLMs can serve as orchestrators or semantic bridges among these components.

    \item \textbf{Low Overhead LLM-enhanced RL model:} 
    LLM latency reaches seconds, exceeding the millisecond-level wireless transmission intervals, and computational costs are higher than traditional RL. 
    To mitigate these issues, strategies include model quantization, pruning, and knowledge distillation. The latter involves using the heavy LLM to train a lightweight model, ensuring the system meets the strict real-time and low-power constraints of wireless networks.

\end{itemize}

\section{Conclusions}
In this paper, we have investigated the advantages of utilizing LLMs-enabled RL in wireless networks. LLMs can enhance RL in wireless networks by serving four key roles: as a feature extractor to translate network environment into state vectors; a reward designer to create dynamic rewards from QoS/QoE goals; a policy interpreter to explain agent decisions; and a decision-maker to convert user intents into control actions.
We then conducted a comprehensive review of existing literature, addressing specific challenges, solutions, and potential obstacles related to the application of LLMs-enabled RL in wireless networks. Then, we have proposed a novel LLMs-enabled multi-agent RL framework for service migration. 
Finally, we have highlighted potential future research directions for the application of LLMs-enabled RL in wireless network scenarios.


\bibliographystyle{ieeetr}
\bibliography{myreference}

@ARTICLE{10819473,
  author={Chen, Yuanbin and Guo, Xufeng and Zhou, Gui and Jin, Shi and Ng, Derrick Wing Kwan and Wang, Zhaocheng},
  journal={IEEE Communications Magazine}, 
  title={Unified Far-Field and Near-Field in Holographic MIMO: A Wavenumber-Domain Perspective}, 
  year={2025},
  volume={63},
  number={1},
  pages={30-36},
  doi={10.1109/MCOM.001.2300845}}

@INPROCEEDINGS{10588921,
  author={Ameur, Mazene and Brik, Bouziane and Ksentini, Adlen},
  booktitle={2024 IEEE 10th International Conference on Network Softwarization (NetSoft)}, 
  title={Leveraging LLMs to eXplain DRL Decisions for Transparent 6G Network Slicing}, 
  year={2024},
  volume={},
  number={},
  pages={204-212},
  doi={10.1109/NetSoft60951.2024.10588921}}

@article{cai2025large,
  title={Large Language Model-enhanced Reinforcement Learning for Low-Altitude Economy Networking},
  author={Cai, Lingyi and Zhang, Ruichen and Zhao, Changyuan and Zhang, Yu and Kang, Jiawen and Niyato, Dusit and Jiang, Tao and Shen, Xuemin},
  journal={arXiv preprint arXiv:2505.21045},
  year={2025}
}

@ARTICLE{9372298,
  author={Feriani, Amal and Hossain, Ekram},
  journal={IEEE Communications Surveys  Tutorials}, 
  title={Single and Multi-Agent Deep Reinforcement Learning for AI-Enabled Wireless Networks: A Tutorial}, 
  year={2021},
  volume={23},
  number={2},
  pages={1226-1252},
  doi={10.1109/COMST.2021.3063822}}

@article{zhang2025embodied,
  title={Embodied AI-Enhanced Vehicular Networks: An Integrated Large Language Models and Reinforcement Learning Method},
  author={Zhang, Ruichen and Zhao, Changyuan and Du, Hongyang and Niyato, Dusit and Wang, Jiacheng and Sawadsitang, Suttinee and Shen, Xuemin and Kim, Dong In},
  journal={arXiv preprint arXiv:2501.01141},
  year={2025}
}

@INPROCEEDINGS{10978505,
  author={Habib, Md Arafat and Iturria Rivera, Pedro Enrique and Ozcan, Yigit and Elsayed, Medhat and Bavand, Majid and Gaigalas, Raimundus and Erol-Kantarci, Melike},
  booktitle={2025 IEEE Wireless Communications and Networking Conference (WCNC)}, 
  title={LLM-Based Intent Processing and Network Optimization Using Attention-Based Hierarchical Reinforcement Learning}, 
  year={2025},
  volume={},
  number={},
  pages={1-6},
  doi={10.1109/WCNC61545.2025.10978505}}

@ARTICLE{10720863,
  author={Zhou, Li and Deng, Xinfeng and Wang, Zhe and Zhang, Xiaoying and Dong, Yanjie and Hu, Xiping and Ning, Zhaolong and Wei, Jibo},
  journal={IEEE Transactions on Cognitive Communications and Networking}, 
  title={Semantic Information Extraction and Multi-Agent Communication Optimization Based on Generative Pre-Trained Transformer}, 
  year={2025},
  volume={11},
  number={2},
  pages={725-737},
  doi={10.1109/TCCN.2024.3482354}}

@ARTICLE{10819462,
  author={Tang, Xuhao and Liu, Fagui and Xu, Dishi and Jiang, Jun and Tang, Quan and Wang, Bin and Wu, Qingbo and Chen, C.L. Philip},
  journal={IEEE Transactions on Consumer Electronics}, 
  title={LLM-Assisted Reinforcement Learning: Leveraging Lightweight Large Language Model Capabilities for Efficient Task Scheduling in Multi-Cloud Environment}, 
  year={2024},
  volume={},
  number={},
  pages={1-1},
  doi={10.1109/TCE.2024.3524612}}

@ARTICLE{10812008,
  author={Chen, Xianfu and Wu, Celimuge and Shen, Yi and Ji, Yusheng and Yoshinaga, Tsutomu and Ni, Qiang and Zarakovitis, Charilaos C. and Zhang, Honggang},
  journal={IEEE Network}, 
  title={Communication and Control Co-Design in 6G: Sequential Decision-Making with LLMs}, 
  year={2024},
  volume={},
  number={},
  pages={1-1},
  doi={10.1109/MNET.2024.3520983}}

@ARTICLE{wang2024LESR,
      title={LLM-Empowered State Representation for Reinforcement Learning}, 
      author={Boyuan Wang and Yun Qu and Yuhang Jiang and Jianzhun Shao and Chang Liu and Wenming Yang and Xiangyang Ji},
      year={2024},
      eprint={2407.13237},
      archivePrefix={arXiv},
      primaryClass={cs.AI},
      journal={https://arxiv.org/abs/2407.13237}, 
}

@ARTICLE{10685369,
  author={Zhou, Hao and Hu, Chengming and Yuan, Ye and Cui, Yufei and Jin, Yili and Chen, Can and Wu, Haolun and Yuan, Dun and Jiang, Li and Wu, Di and Liu, Xue and Zhang, Charlie and Wang, Xianbin and Liu, Jiangchuan},
  journal={IEEE Communications Surveys  Tutorials}, 
  title={Large Language Model (LLM) for Telecommunications: A Comprehensive Survey on Principles, Key Techniques, and Opportunities}, 
  year={2024},
  volume={},
  number={},
  pages={1-1},
  doi={10.1109/COMST.2024.3465447}}

@ARTICLE{10375570,
  author={Wang, Guanhua and Yang, Fang and Song, Jian and Han, Zhu},
  journal={IEEE Transactions on Communications}, 
  title={Optimization for Dynamic Laser Inter-Satellite Link Scheduling With Routing: A Multi-Agent Deep Reinforcement Learning Approach}, 
  year={2024},
  volume={72},
  number={5},
  pages={2762-2778},
  doi={10.1109/TCOMM.2023.3347775}}

@ARTICLE{10766898,
  author={Cao, Yuji and Zhao, Huan and Cheng, Yuheng and Shu, Ting and Chen, Yue and Liu, Guolong and Liang, Gaoqi and Zhao, Junhua and Yan, Jinyue and Li, Yun},
  journal={IEEE Transactions on Neural Networks and Learning Systems}, 
  title={Survey on Large Language Model-Enhanced Reinforcement Learning: Concept, Taxonomy, and Methods}, 
  year={2024},
  volume={},
  number={},
  pages={1-21},
  doi={10.1109/TNNLS.2024.3497992}}

@ARTICLE{10742906,
  author={Park, Jihong and Ko, Seung-Woo and Choi, Jinho and Kim, Seong-Lyun and Choi, Junil and Bennis, Mehdi},
  journal={IEEE BITS the Information Theory Magazine}, 
  title={Toward Semantic MAC Protocols for 6G: From Protocol Learning to Language-Oriented Approaches}, 
  year={2024},
  volume={4},
  number={1},
  pages={59-72},
  doi={10.1109/MBITS.2024.3491949}}

@ARTICLE{10839354,
  author={Yang, Tingting and Feng, Ping and Guo, Qixin and Zhang, Jindi and Ning, Jiahong and Wang, Xinghan and Mao, Zhongyang},
  journal={IEEE Transactions on Cognitive Communications and Networking}, 
  title={AutoHMA-LLM: Efficient Task Coordination and Execution in Heterogeneous Multi-Agent Systems Using Hybrid Large Language Models}, 
  year={2025},
  volume={},
  number={},
  pages={1-1},
  doi={10.1109/TCCN.2025.3528892}}
\vspace{-18mm}
\begin{IEEEbiographynophoto}
{Jie Zheng} received 
the PhD degree from the Department of Telecommunications Engineering, Xidian University, China. He is currently a full professor with the College of Computer Science, Northwest University, Xi’an, China. 
\end{IEEEbiographynophoto}
\vspace{-18mm}
\begin{IEEEbiographynophoto}
{Ruichen Zhang} received the Ph.D. degree in Beijing Jiaotong University, Beijing, China. He is the postdoctoral research fellow in the College of Computing and Data Science, Nanyang Technological University, Singapore. 
\end{IEEEbiographynophoto}
\vspace{-18mm}
\begin{IEEEbiographynophoto}
{Dusit Niyato}
received 
the PhD degree in electrical and computer engineering from the University of Manitoba, Canada. He is currently a full professor with the School of Computer Science and Engineering, Nanyang Technological University, Singapore. 
\end{IEEEbiographynophoto}
\vspace{-18mm}
\begin{IEEEbiographynophoto}
{Haijun Zhang} was a Post-Doctoral Research Fellow with the Department of Electrical and Computer Engineering, The University of British Columbia (UBC), Canada. He is currently a Full Professor with the University of Science and Technology Beijing, China. 
\end{IEEEbiographynophoto}
\vspace{-18mm}
\begin{IEEEbiographynophoto}
{Jiacheng Wang} received the Ph.D. degree Chongqing University of Posts and Telecommunications, Chongqing, China. He is the postdoctoral research fellow in the College of Computing and Data Science, Nanyang Technological University, Singapore. 
\end{IEEEbiographynophoto}\vspace{-18mm}
\begin{IEEEbiographynophoto}
{Hongyang Du} 
received the PhD degree with at Nanyang Technological University, Singapore.  He is currently an assistant professor with University of Hong Kong, China. 
\end{IEEEbiographynophoto}

\vspace{-18mm}
\begin{IEEEbiographynophoto}
{Jiawen Kang}
received the PhD degree from the Guangdong University of Technology, China. He was a post-doctoral researcher with Nanyang Technological University, Singapore. He is currently a full professor with the Guangdong University of Technology. 
\end{IEEEbiographynophoto}
\vspace{-18mm}
\begin{IEEEbiographynophoto}
{Zehui Xiong}
received 
the Ph.D. degree in computer science and engineering from Nanyang Technological University (NTU), Singapore. He is currently a full professor with Queen’s University Belfast, United
Kingdom. 
\end{IEEEbiographynophoto} 

\end{document}